\documentclass[namedreferences]{solarphysics}
\usepackage[hyperref,optionalrh,solaromanenum]{spr-sola-addons} 
\usepackage{graphicx}                    
\usepackage{color}                       

\usepackage{amssymb}
\usepackage{latexsym}
\usepackage{rotating}
\usepackage{float}
\usepackage{booktabs}
\usepackage{multirow}
\usepackage{makecell}

\usepackage{url}
\usepackage{xcolor}
\definecolor{newcolor}{rgb}{.8,.349,.1}
\usepackage{changepage,threeparttable}
\usepackage {multicol}
\usepackage{caption}		
\usepackage{txfonts}
\usepackage{longtable}
\usepackage{tabularx}
\usepackage{rotating}
\usepackage{pdflscape}
\usepackage{soul} 
\newcommand{\rs}[1]{R$_\odot$}

\begin{document}

\begin{article}

\begin{opening}

\title{Kilometric type II radio emissions in Wind/WAVES TNR data and association with interplanetary structures near Earth}

%
\author[addressref={1},corref,email={franco.manini@um.edu.ar}]{\inits{F. }\fnm{Franco }\lnm{Manini}\orcid{0000-0002-8455-2632}}
\author[addressref={1},corref]{\inits{H. }\fnm{Hebe }\lnm{Cremades}\orcid{0000-0001-7080-2664}}
\author[addressref={1},corref]{\inits{F.M. }\fnm{Fernando M. }\lnm{López}\orcid{0000-0002-2047-6327}}
\author[addressref={2},corref]{\inits{T. }\fnm{Teresa }\lnm{Nieves-Chinchilla}\orcid{0000-0003-0565-4890}}

%
\runningauthor{Franco Manini}
\runningtitle{Kilometric type II radio emissions in Wind/WAVES TNR data}

\address[id={1}]{Universidad de Mendoza, CONICET, Grupo de Estudios en Heliofisica de Mendoza, Boulogne Sur Mer 665 (5\,500) Mendoza, Argentina}
\address[id={2}]{Heliospheric Physics Laboratory, Heliophysics Science Division, NASA Goddard Space Flight Center, 8800 Greenbelt Rd., Greenbelt, MD 20770, USA}

\begin{abstract}
Type II radio bursts arise as a consequence of shocks typically driven by coronal mass ejections (CMEs). When these shocks propagate outward from the Sun, 
their associated radio emissions drift down in frequency as excited particles emit at the local plasma frequency, creating the usual Type II patterns. In this work, we use  dynamic spectra from the Wind/WAVES Thermal Noise Receiver (TNR) to identify Type II radio emissions in the kilometric wavelength range (kmTIIs, $f <$ 300 kHz) between 1 January 2000 and 31 December 2012, i.e. over a solar cycle. We identified 134 kmTII events and compiled various characteristics for each of them. Of particular importance is the finding of 45 kmTII events not reported by the official Wind/WAVES catalog (based on RAD1 and RAD2 data). We search for associations with interplanetary structures and analyze their main characteristics, in order to reveal distinctive attributes that may correlate with the occurrence of kmTII emission. We find that the fraction of interplanetary coronal mass ejections (ICMEs) classified as magnetic clouds (MCs) that are associated with kmTIIs is roughly similar to that of MCs not associated with kmTIIs. 
Conversely, the fraction of ICMEs with bidirectional electrons is significantly larger for those ICMEs associated with kmTIIs (74\,$\%$ vs. 48\,$\%$). Likewise, ICMEs associated with kmTIIs are on average 23\,$\%$ faster. The disturbance storm time (DsT) mean value is almost twice as large for kmTII-associated ICMEs, indicating that they tend to produce intense geomagnetic storms. In addition, the proportion of ICMEs producing moderate to intense geomagnetic storms is twice as large for the kmTII-associated ICMEs. After this investigation, TNR data prove to be valuable not only as complementary data for the analysis of kmTII events but also for forecasting the arrival of shocks at Earth.

\end{abstract}

%

\end{opening}

%
\section{Introduction}\label{sec:intro}

Observations in the radio frequency domain provide an important source of information about physical processes that take place in the solar corona and the heliosphere. Of particular interest are the radio observations in the low frequency range, from metric (300-30 MHz) to kilometric wavelengths (f $<$ 300 kHz). It is in this frequency interval where it is possible to observe type II radio bursts \citep{wild1950,wild1954}. 
Radio emissions associated to type II bursts are believed to originate in the upstream regions of magneto-hydrodynamic (MHD) shocks mostly driven by coronal mass ejections (CMEs) propagating outward in the solar corona and the interplanetary medium \citep{reiner97,bale99}. 

During its propagation in the heliosphere, a CME-driven shock encounters a plasma density that decreases with increasing heliocentric distance. As the type II emission frequency is directly related with the local plasma density, the outward-propagating shock gives rise to an emission pattern that drifts down to lower frequencies as time (and distance) progresses \citep{reiner97}. The frequency of radio bursts can give a rough estimation of the heliocentric distance $(R)$ at which it is produced \citep[e.g.][]{Leblanc-etal1998}. Although the generation mechanism of Type II bursts is to date not fully understood, they can be regarded as electrons undergoing shock-drift acceleration and developing a beam distribution of reflected electrons at shocks. This enables growth of Langmuir waves via the beam instability in the upstream foreshock, while Langmuir wave energy is converted into radio emission near the electron plasma frequency and its harmonic, via nonlinear wave-wave processes \citep{knock01, chernov21}. Dynamic spectra (DS) thus constitute a valuable resource to visualize these drifting emissions among other radio phenomena \citep{reiner98a}. 

Observations below 300\,kHz can be used to detect the signal from type II bursts emitted at kilometric wavelengths (hereafter kmTII), which correspond to distances roughly beyond 20 solar radii (\rs{}) and all the way down to 1\,AU. Particularly for the case of long-wavelength Type IIs, it is worth mentioning that the fact that they extend to such low frequencies is indicative that the shock must still be highly energetic at large distances to keep the plasma emission mechanism energized, turning kmTIIs into a valuable and special asset. Also, according to \cite{corona15}, all the
interplanetary shock waves associated with kmTIIs are explosive beyond 20 \rs{}.

The first report of a kmTII was made in 1973 \citep{malitson1973}. Later, in 1982, \cite{cane82} identified several kmTII events using data from ISEE-3. \cite{cane87} found that kmTII bursts were associated with the most energetic and massive CMEs observed by the Solwind coronagraph.
A major advance in radio observations at low frequencies came with the launch of the Wind mission, in particular with the instrument Radio and Plasma Wave Experiment \citep[WAVES, ][]{Bougeret95}. The use of data provided by WAVES has allowed the analysis of plasma emission from type II bursts from the solar corona to the Earth environment \citep[e.g., ][]{Cane05, gonzalez-esparza09,Hillan12,Xie12}.
More specifically, the Wind/WAVES receiver RAD2 (Radio Receiver Band 2; 1-14 MHz) can be used for the detection of type II events in the metric-decametric wavelength range, while  RAD1 (Radio Receiver Band 1; 20-1040 kHz) can be used for those in the hectometric-kilometric range.
Such list of events detected by these two receivers has been compiled and made available online in the Wind/WAVES official website\footnote{\url{https://solar-radio.gsfc.nasa.gov/wind/data_products.html}} as a ``Possible Type II and IV Radio Bursts Observed by WIND/WAVES'' (hereafter ``Wind list''). RAD1 and RAD2 data have been intensively used to analyze type II radio bursts \citep[e.g., ][]{kaiser1998, dulk99, Reiner_2007, Gopal19}. 

The Thermal Noise Receiver (TNR; 4-256 kHz), is also part of the Wind/WAVES instrument suite. Efforts dealing with TNR data \citep[e.g., ][]{gonzalez-esparza, Aguilar-Rodriguez} are scarce in comparison to those using the other Wind/WAVES detectors.
Although RAD1 and TNR may appear to largely overlap in frequency, the latter has a much higher spectral resolution, which turns it into a valuable asset for the detection and analysis of kmTII events. Taking advantage of TNR's capabilities, \cite{cremades07} developed a technique to estimate the speed of shocks based on the associated kmTII emission. In a similar way, \cite{cremades15} analyzed 71 Earth-directed shocks driven by CMEs to predict their arrival time at 1\,AU. These papers remark the importance of using kmTII emission for tracking CMEs in the interplanetary medium (ICMEs), a fundamental aspect in space weather forecasting.  
Nonetheless, a complete survey of kmTII radio emissions identified in the Wind/WAVES TNR data and their association with interplanetary (IP) structures has not been carried out to date. This is a crucial step towards understanding which characteristics of IP structures are prone to result in kmTII emissions. Moreover, the geoeffectiveness of IP structures associated to kmTII emissions can be assessed to acknowledge for differences with respect to those not associated to kmTII emissions.
Through the analysis of TNR DS, here we present the results of a careful survey in which we compiled a list of 134 kmTII events. These events were subsequently associated with interplanetary transient events detected in-situ at Earth-L1, such as shocks and ICMEs.

The manuscript is organized as follows. In Section \ref{sec:identif} we 
describe the methodology used to identify the kmTII events. In Section \ref{sec:assoc_inter} we explain the procedure followed to associate the IP structures detected in situ at 1\,AU with our kmTII events. 
In Section \ref{sec:statistics} we summarize the results concerning statistics, characteristics of associated ICMEs and geoeffectiveness. Finally, in Section \ref{sec:conclusions} we summarize our findings and discuss their implications. The complete table of identified events can be found in the Appendix at the end of the manuscript.

%

%

\section{Identification of the kmTII events}
\label{sec:identif}

In this section we present the procedure used to identify the kmTII events using data from the Wind/WAVES TNR radio receiver, and the compiled list of events.
The access to the Wind data is open and performed through the Wind/WAVES official website as mentioned above.  

\begin{figure*}[htb]
\centering
\includegraphics[trim={0.1cm 2cm 0cm 2.5cm},clip, width=1.0\textwidth]{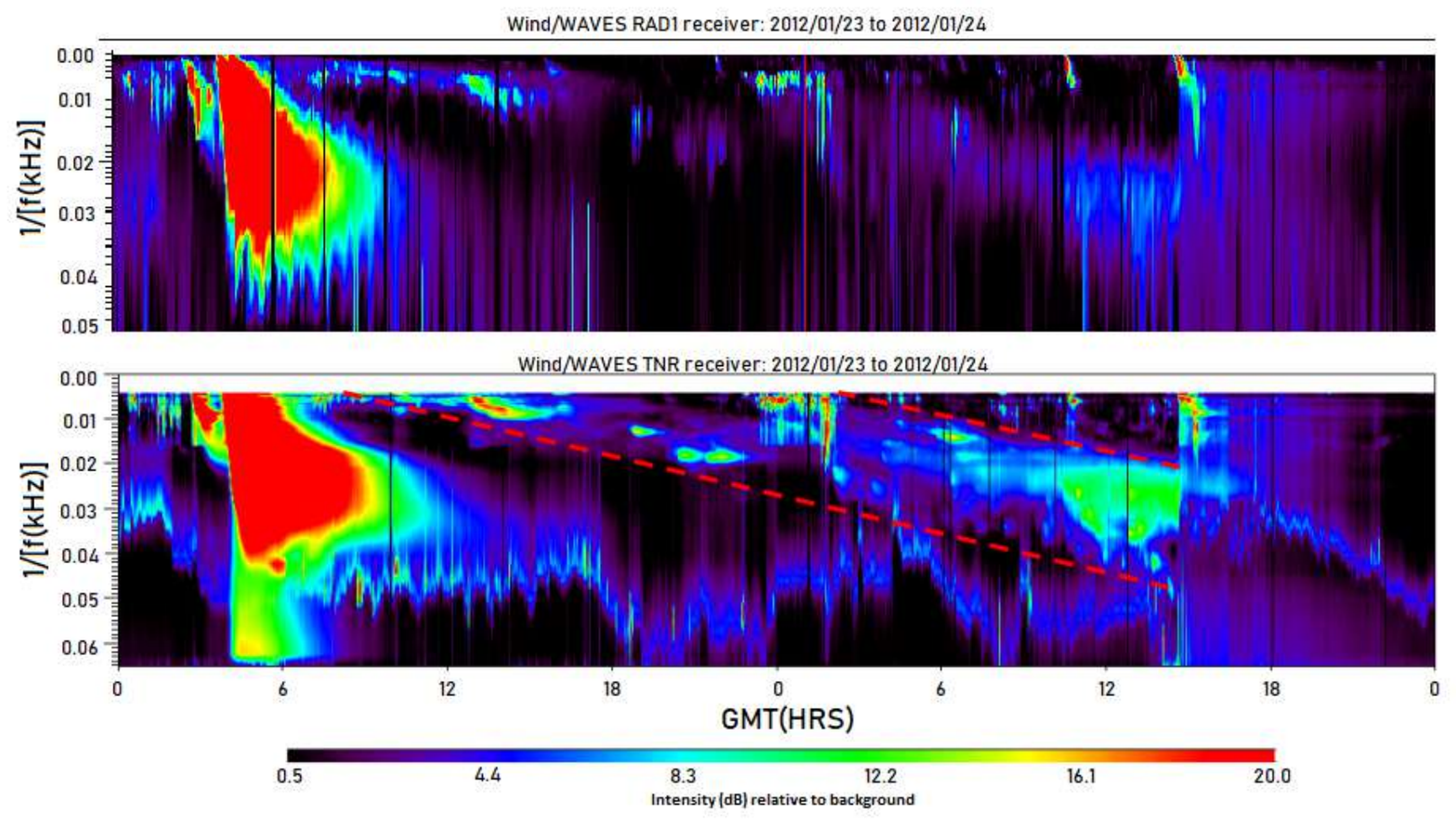}
\caption{DS showing the radio emission detected by RAD1 (upper panel) and TNR (lower panel) during 23 and 24 of January 2012. The emission between the red dashed lines in the lower panel is a clear kmTII event observed during those days, that has also left an imprint in the RAD1 DS. The upper panel covers the range of frequencies 1040-20 kHz, and the lower panel 256-15 kHz.}

\label{Fig1}
\end{figure*}

To identify the kmTII events, we generate DS from the signal detected by TNR. For the purpose of this analysis, we built DS that display the radio emission in the inverse frequency domain ($f^{-1}$) as a function of time (see sample DS in Figure \ref{Fig1}). We prefer to use $f^{-1}$ because the average IP plasma density varies as $1/R^2$, and the plasma frequency $f\,[kHz]$ is proportional to the electron density $n_e [\,\mathrm{cm}^{3}]$ by $f \propto \sqrt{n_e }$, the plasma frequency scales as $1/R$, thus $1/f$ scales as $R$ \citep{reiner97}. An emission observed in a DS is considered as a kmTII event if the two following conditions are met: i) it shows a substantial decay in frequency, of at least 0.5\,kHz per hour (In the $1/f$ domain, this approximately equals to 0.00133\,kHz$^{-1}$ per hour) and ii) the emission is observed in the DS 
in connection with an eruption previously occurred, related to its origin in the low solar corona. A typical signature associated with an eruption is a Type III radio burst, but if the latter is not present, the eruption can be identified in e.g. Extreme Ultraviolet (EUV) imagery of the low corona.

In Figure \ref{Fig1} we show a DS that exemplifies the conditions described above. The figure displays a case of type II radio emission occurred during the rising phase of solar cycle 24, with the upper panel corresponding to RAD1 frequencies and the lower panel to TNR. Type II radio emission can be discerned in both DS, with an improved resolution in frequency for the case of TNR. In the TNR panel, the kmTII is delimited by the red dashed lines that meets both aforementioned conditions. A type III radio burst can be seen in both DS, in association with the start of the event.
In the lower frequency region of the lower panel, it is possible to observe the plasma frequency line (PFL) as a wandering continuous emission at all times.
A shock wave can also be noticed where the red dashed lines end, at 14:40 GMT on 24 January. The shock arrival produces a sudden increase in the frequency of the emission associated with the PFL, due to the shock density enhancement. 

Our investigation uses 12 years of data, from 1 January 2000 to 31 December 2012, thus partially covering solar cycles 23 and 24. From the analysis of all DS during that time interval, we could identify 134 kmTII events. All 134 compiled events are listed in Table \ref{table:associations}, together with their main characteristics and associated interplanetary structures. While the radio events themselves are introduced in this section, the properties of the associated shocks and ICMEs are described in Section \ref{sec:assoc_inter}. The list is also available for download at \url{https://sites.google.com/um.edu.ar/gehme/science/km-tii-catalogue}.

The first group of columns 'KmTII' (1-5) in the table displays information on the identified kmTII events: start date and time, end date and time, and frequency range in kHz.
It is important to remark that we could not find any kmTII event during the years 2008, 2009 and 2010.  The column 'Wind/WAVES' (6) in Table \ref{table:associations} lists the related events that are reported by the Wind list using the same reference number of the event as listed, following the corresponding year. Those events without reference number in this column, are kmTII events identified by us, that are not reported by the Wind list.
The procedure of association of each identified kmTII in the DS of TNR with a counterpart registered in the official Wind list, required two conditions that must be simultaneously met: 
i) There must be a temporal agreement between both entries 
and ii) 
The lower limit in frequency of the emission reported by the Wind list should extend beyond the upper limit of TNR's frequency detection range.
It is worth noting that there are events which had been catalogued in the Wind list as reaching down to TNR frequencies but not observed in the DS of TNR, therefore not considered in our analysis.

The second group of columns 'In situ shock' displays information on the association with shocks detected by the Wind and/or ACE spacecraft. The third group of columns 'ICME' summarizes the main characteristics of the associated ICMEs. These two groups of columns will be addressed in detail in Section \ref{sec:assoc_inter}.



\section{Association between kmTII and interplanetary transient phenomena detected in-situ at 1\,AU} \label{sec:assoc_inter}

Once the kmTII events are identified in the DS of TNR, we searched for their spatial and temporal association with interplanetary transient structures detected in-situ at the Earth's environment. To accomplish this task we used data available from several online catalogs, as described below. 

In the case of interplanetary shocks, we used the database compiled by the Harvard-Smithsonian Center for Astrophysics (CfA\footnote{\url{https://www.cfa.harvard.edu/shocks/}}) which is based in data from the Wind and the Advanced Composition Explorer (ACE) spacecraft. Alternatively, we used the ``List of disturbances and transients'' compiled by the University of New Hampshire (UNH\footnote{\url{http://www.ssg.sr.unh.edu/mag/ace/ACElists/obs_list.html}}).

As mentioned above, kmTIIs typically correspond to distances beyond 20\,\rs{} and all the way down to 1\,AU. The average starting frequency of the kmTIIs identified in this study is 185 kHz ($\approx$ 33\,\rs{}), while they typically end at 54 kHz ($\approx$ 117\,\rs{}). This poses them as a valuable resource to predict shock arrival times in advance.
Having said this, to associate a kmTII to an interplanetary shock, we considered as possible candidates those shocks detected no more than four days since the beginning of the kmTII emission.
This criterion is an upper limit assuming that a CME traveling at 400\,km s$^{-1}$ would take approximately that amount of time to reach 1\,AU. If more than one shock is registered within such time interval, we examined the DS following the procedure described in \cite{cremades07}: to find the most likely associated shock, we used a linear fit to extrapolate forward in time the kmTII emission which should intersect the abrupt signal increment of the PFL given by the arrival of the respective shock at 1\,AU. Then, the intersection time of the fit is compared with the in-situ arrival time provided by the shock catalogs. The CfA Wind shock catalog was always consulted first. In those cases where we could not associate any shock reported there, we conducted the search in the CfA ACE and UNH shock databases.

The result of following this association procedure is displayed in the second group of columns in Table \ref{table:associations} (In situ shock). Columns 7 to 10 display information about the start date and time for the shocks locally detected by the Wind and/or ACE spacecraft, median shock speed, and the catalog used to extract that information. The blank spaces indicate that we could not associate any shock to the corresponding kmTII event. In this respect, it is worth noting that even though we assume for a first approximation that observing a kmTII with Wind/WAVES implies that the shock/CME is Earth directed, this might not necessarily be the case. These radio emissions are widely beamed, thus they can be detected at a particular location (i.e. Wind) without implying that the shock is actually traveling in that direction. That is, the solid angle through which kmTII associated shocks/CMEs can be emitted such that they intersect L1 is smaller than the solid angle through which they can be emitted and their radio emission be detected at L1.
Therefore, it is highly likely that for several detected kmTII, we will not be able to observe their associated shocks. See section \ref{sec:conclusions} for additional discussion regarding this subject.

To study the association between the identified kmTII events and the observed ICMEs, we use the following catalogs of ICME transient events: The Richardson-Cane list \citep[][hereafter RC\footnote{\url{http://www.srl.caltech.edu/ACE/ASC/DATA/level3/icmetable2.html}}]{RC03,RC10}, using the ACE spacecraft, and the one compiled by Teresa Nieves-Chinchilla \citep[][hereafter TNC\footnote{\url{https://wind.nasa.gov/ICME_catalog/ICME_catalog_viewer.php}}]{NC18, NC19}, using the Wind spacecraft.

In our analysis, we assume that an ICME is related to the kmTII, if such ICME starts right after a cataloged interplanetary shock, and/or within a four-day period since the kmTII start. For those cases without shock association, the kmTII profiles were linearly projected down to the PFL to have a proxy of the date and time of arrival, so as to enable an ICME association in RC and/or TNC catalogs.

The third group of columns (11-19) in Table \ref{table:associations} summarize the main characteristics of the associated ICMEs. Columns 11 and 12 identify the start date and time of the ICMEs that we could associate with our kmTII events. Column 13 (``MC'') indicates the classification used in RC \citep[described in][]{RC03,RC10} with regards to a magnetic cloud (MC): ``0'' means no detection of MC, ``1'' some characteristics are present but lacks rotation or enhanced magnetic field, and ``2'' all MC requirements are fulfilled. Column 14 (``MO'', for magnetic obstacle) indicates the type of magnetic structure according to TNC \citep[as described in][]{NC18,NC19}: ``Fr'' flux rope, ``F+'' or ``F-'' flux rope with large or small rotation respectively, ``Cx'' Complex, and ``E'' Ejecta. Column 15 (``BDE'') indicates the presence of bidirectional electrons in ACE/SWEPAM observations according to RC. When 'SEP' appears, it means that a storm was in progress so it could not be determined. Column 16 (V$_{\mathrm{ICME}}$) presents the mean ICME speed, based on solar wind speed observations during the period of the passage of the ICME, in the RC catalogue. For those events catalogued only in the TNC catalogue, this value corresponds to the mean solar wind bulk velocity (V$_{sw}$ in that table) in the Magnetic Object interval. Column 17 (``B'') is the mean magnetic field strength in the ICME. When only the TNC catalog is used, the value is for the MO interval. Column 18 shows the disturbance storm time index (DsT) associated with the corresponding ICMEs after their arrival to Earth. For those cases where the corresponding ICMEs were only listed in the TNC catalog, we estimated the corresponding DsT values manually\footnote{As a consequence of this work, we provided the TNC catalog with DsT values for all their entries. See acknowledgments in \url{https://wind.nasa.gov/ICMEindex.php}.}. To this end, we followed the same criteria used in the RC list, namely taking the minimum value reported by the World Data Center for Geomagnetism, Kyoto\footnote{\url{https://wdc.kugi.kyoto-u.ac.jp/dst_final/index.html}} during the ICME duration interval. The last column, 19, indicates the catalog (RC and/or TNC) used to extract the ICMEs information in every case. Blank spaces indicate that no ICME from the catalogs could be associated with the respective kmTII event.

\section{Results}
\label{sec:statistics}

\subsection{Statistical analysis of the identified events}

A summary of the resulting associations between kmTII events and interplanetary structures detected in situ is displayed in Table \ref{table:resumen}. 
Based on the analysis of TNR data, we could identify 134 kmTII radio emission events, out of which 45 (33$\%$) are not listed in the Wind's list, i.e. these events were not reported on the basis of data from RAD1 and/or RAD2.  

From the total number of events, 77 (57$\%$) could be associated to an interplanetary shock, hence 57 kmTII emission events were not. It is worth noting that for the 89 kmTIIs that were reported by the Wind's list, 54 (61$\%$) can be associated with interplanetary shocks, while this number is 23 (51$\%$) for those not previously reported.  Furthermore, 62 events (47\%) out of the total amount of kmTIIs could be associated with an ICME, while the proportion is practically the same for the subsets of events previously reported by the Wind's list and those that were newly identified in this work. 
In the same way, the relationships between the amount of events with ICMEs and MCs, with and without shockwaves, is similar between our list and the Wind list. 

 \begin{table}[htbp]
 \caption{Number of identified kmTII events and their association with in-situ interplanetary structures. }
\captionsetup{width=1\textwidth}
\begin{tabular}{cccc}
\hline
  & All & Prev. reported & Newly reported  \\
\hline
 KmTII events &  134 & 89 & 45 \\
 With shockwave & 77 & 54 & 23 \\
 With ICME & 62 & 42 & 20\\
 With shock and ICME & 51 & 35 & 16 \\
 With shock - No ICME & 26 & 19 & 7 \\
 With ICME - No shock & 11 & 6 & 5 \\
 \hline
\end{tabular}
\label{table:resumen}
\end{table}

\begin{figure*}[htb]
\centering

\includegraphics[width = 0.9\textwidth]{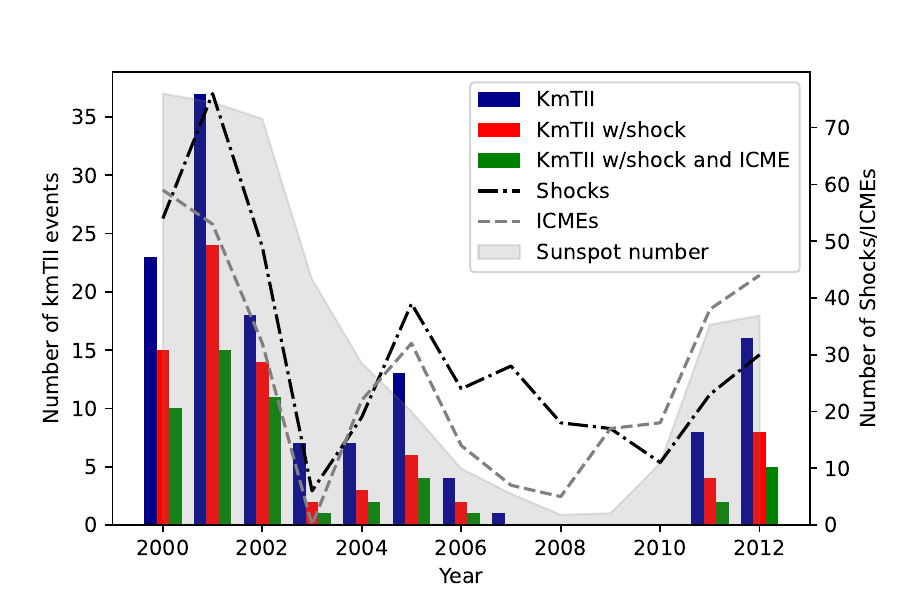}
\caption{Yearly frequency of identified kmTIIs (blue), kmTIIs with associated shocks (red), and kmTIIs associated with both, a shock and an ICME (green). The yearly number of shocks and ICMEs are shown in black and grey lines respectively. The scales of the primary and secondary y-axes differ by a factor of 2, given that the number of shocks/ICMEs almost double the number of kmTIIs. The gray shaded area represents the yearly sunspot number.}
\label{fig:shocks_hist}
\end{figure*}

To examine the variation with the solar cycle, we display in Figure \ref{fig:shocks_hist} the total number of kmTII events compiled per year, as well as how many of them could be associated to an interplanetary shock and to both, a shock and an ICME. 
The figure indicates that for years 2000 to 2002 the ratio of kmTII events with associated shock waves is significantly larger than 50\% with respect to the total amount of kmTII events, thus corresponding to the period of maximum activity of solar cycle 23. For year 2003, this ratio is considerably lower, while for years 2004\,--\,2006 and 2011\,--\,2012 the proportion is around 50\%. Besides, when considering the number of kmTII events associated with both, shocks and ICMEs, the ratio with respect to the total of kmTII events is lower than 50\% for every year, except for 2002 when it is significantly larger. The fact that this fraction is not roughly constant over the solar cycle, as it would be expected if only the larger beam width vs. the shock extent (Section \ref{sec:assoc_inter}) is responsible for the difference, suggests that other mechanisms might be at work (see Section \ref{sec:conclusions}).     
The yearly number of all shocks reported by Wind and ACE, and all ICMEs reported by RC and TNC are shown for reference in dashed-dotted and dashed lines respectively. The yearly number of shocks (from both CfA Wind and ACE catalogs) arises from considering all shock events, counting only once those that are duplicated in the catalogs. The same procedure was applied to compile the yearly number of ICMEs, arising from the RC and TNC catalogs.

\subsection{Comparison of ICMEs with and without associated kmTII}
\label{sec:comparison}

In view to understand if there are intrinsic differences between ICMEs associated with kmTII (kmTII-ICMEs) and those that are not (other-ICMEs), here we compare the properties of both ICME groups. As mentioned in Table \ref{table:resumen}, the kmTII-ICMEs group is comprised of 62 events. The procedure to compile the other-ICMEs group is as follows:
As a starting point, we compiled a joint list of ICMEs from both TNC and RC catalogues during the investigated time period (2000-2012), and removed those entries associated with kmTIIs. Every event was carefully checked to have been computed only once in the merged other-ICMEs list, which is finally comprised of 302 cases. 

To characterize both of the ICME groups, we consider the following aspects: MC, BDE and V$_{\mathrm{ICME}}$. These properties were described in Section \ref{sec:assoc_inter}, while the respective values corresponding to the group kmTII-ICMEs are presented in columns 13, 14, 15 and 16 of Table \ref{table:associations}. With regards to the MC aspect, it becomes necessary to reconcile the classification used by the TNC and RC catalogues. We associate the TNC classification (E, Cx, F; Table \ref{table:associations}, column 14) with that of RC (0, 1, 2; Table \ref{table:associations}, column 13), by assuming E=0, Cx=1 and F=2. The results of this analysis are summarized in Table \ref{table:insitu}. 

Our study indicates that the fraction of ICMEs classified as MCs is quite similar for the cases where the events were and were not associated with kmTII emission, resulting in 21 out of 62 (34\%) and 118 out of 302 (39\%) events respectively. 

The comparison of the detection of BDEs does show a remarkable difference between both groups (see second row of Table \ref{table:insitu}). For the kmTII-ICMEs group, the number of events which present BDEs is of 46 out of 62 (74\%), while for the other-ICMEs group the proportion is 145 out of 302 (48\%), which is significantly lower. 

\begin{figure*}[ht]
\centering
\includegraphics[width = 0.95\textwidth]{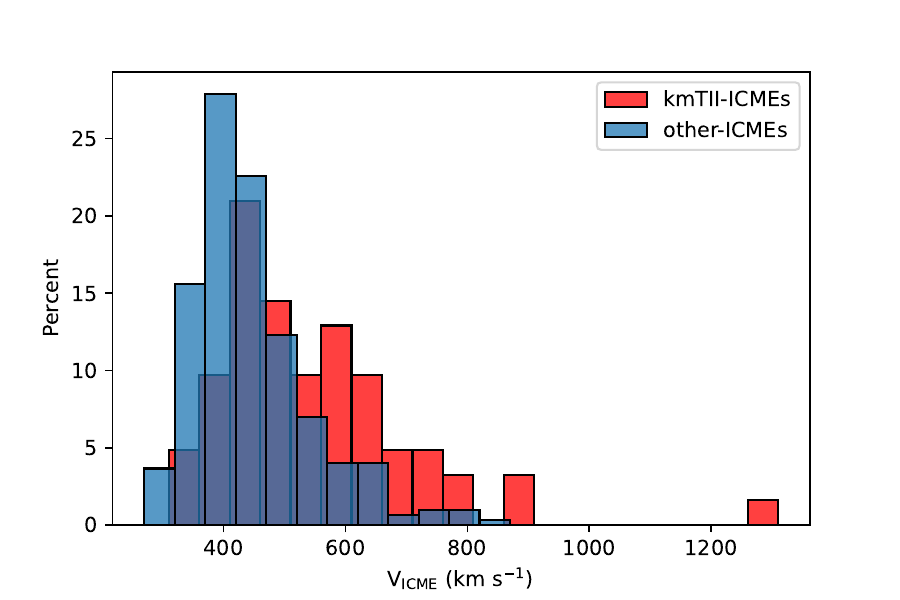}
\caption{Histograms of (V$_{\mathrm{ICME}}$) for both groups of analysis. Given the difference in number of events, a percent normalization was used, such that bar heights sum to 100.
}
\label{fig:vel_max}
\end{figure*}

In Figure \ref{fig:vel_max} we show histograms of V$_{\mathrm{ICME}}$ for both analyzed groups. The mean values of V$_{\mathrm{ICME}}$ in Table \ref{table:insitu} indicate that on average, the events of the kmTII-ICMEs group are around 23\% faster than those without evidence of kmTII. The median values are 515\,km s$^{-1}$ and 420\,km s$^{-1}$ respectively, also supporting the large difference between both groups. 
This result is in agreement with the recent findings of \citet{Patel2022}. They compare the mean values of V$_{\mathrm{ICME}}$ for ICMEs detected during solar cycles 23 and 24, associated and not associated with decametric and hectometric type II radio emission, finding similar values to those of our study. Also, \cite{Gopalswamy08} found that metric and kilometric type II bursts are associated with CMEs with higher average speeds.

\begin{table}[htbp]
\caption{Proportion of ICMEs for the two considered groups showing characteristics of MC and BDE. The last row corresponds to the arithmetic mean of V$_{\mathrm{ICME}}$, with the standard deviation as uncertainty. }
\begin{tabular}{ccc}
\toprule
 & kmTII-ICMEs & other-ICMEs  \\
\midrule
MC (\%) & 34 & 39 \\
BDE (\%) & 74 & 48 \\
$\bar{V}$$_{\mathrm{ICME}}$ [km\,s$^{-1}$] & 545 $\pm$ 163 & 442 $\pm$ 97 \\ 
\bottomrule\label{table:insitu}
\end{tabular}
\end{table}

\subsection{Geoeffectiveness of ICMEs with and without associated kmTII}
\label{sec:geoeffectiveness}

Of particular interest is the analysis of the geoeffectiveness resulting from the arrival of ICMEs which present kmTII emission and from those that do not. With this objective, we consider the same groups, kmTII-ICMEs and other-ICMEs, introduced in Section \ref{sec:comparison}. 

In order to quantify geoeffectiveness we use the DsT index (see Section \ref{sec:assoc_inter}). The mean Dst for the kmTII-ICMEs group yields -111, while that of the other-ICMEs group is -53. This result indicates that, if considering the mean values, the DsT of those ICMEs associated with kmTII emission is twice as large (in negative values) than that of those ICMEs not associated with kmTII.  

\begin{figure*}[htb]
\centering
\includegraphics[width = 0.95\textwidth]{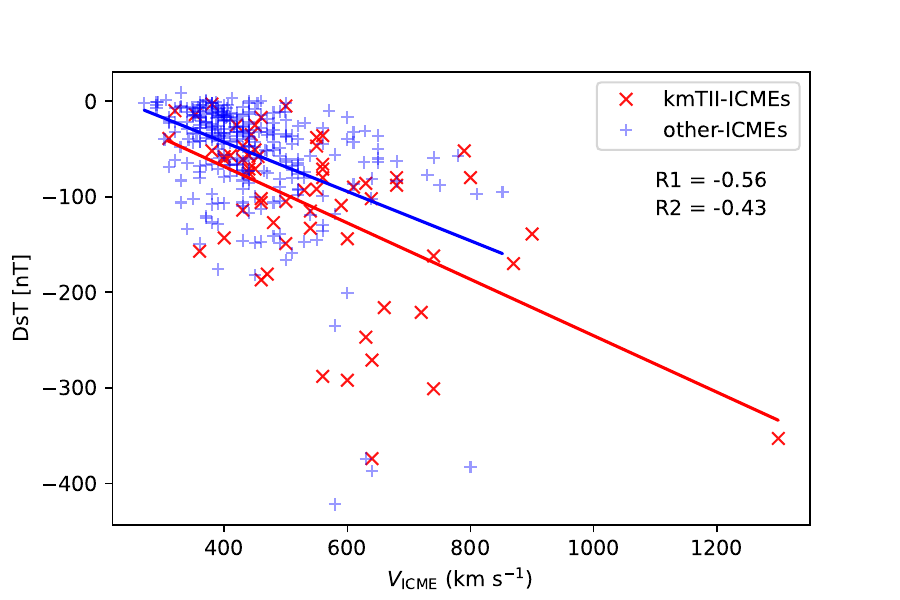}
\caption{DsT values vs. V$_{\mathrm{ICME}}$, for all considered events in the kmTII-ICMEs and other-ICMEs groups. Linear fits are displayed in the same color than its group. The linear correlation coefficient R1 corresponds to kmTII-ICMEs and R2 to other-ICMEs.}
\label{fig:vel_dst}
\end{figure*}

In Figure \ref{fig:vel_dst} we present a plot of DsT versus the mean speed of the associated ICMEs. The figure suggests that for both considered groups, faster ICMEs are related with the occurrence of stronger geomagnetic storms \citep[characterized by higher DsT in absolute value;][]{Gopalswamy09}. However, the linear correlation between DsT and V$_{\mathrm{ICME}}$ is weak, on the basis of the correlation coefficients for both data series.

To specifically analyze those ICMEs associated to moderate and intense geomagnetic storms \citep{Gonzalez99}, we considered only those events whose DsT\,$\leq$\,-50 \citep{Loewe97}. 
With this consideration, we found that out of the 62 events of the group kmTII-ICMEs, 49 (79\%) produced moderate to intense geomagnetic storms. These events present a mean DsT value of -133, with a median of -105. The other-ICMEs group, originally comprised of 302 events, was left with 115 (38\%) after removing those events with DsT\,$>$\,-50. For this group we found a mean DsT value of -102, with a median of -81. These results suggest that kmTII-related ICMEs are more prone to produce larger geomagnetic storms. In fact, these are up to 25\% larger than for those ICMEs not linked to kmTII emission events. This is in agreement with \cite{Patel2022}, who also found that decametric and hectometric type II-related ICMEs produced more intense geomagnetic storms, with faster ICMEs on average. Moreover, we also find that the proportion of ICMEs producing moderate to intense geomagnetic storms is twice as large for the kmTII-ICMEs group than for the other-ICMEs group (79\% vs. 38\% respectively).

\section{Discussion and conclusions}\label{sec:conclusions}

According to the identification criteria adopted in this work, we compiled a list of 134 kmTII events on the basis of TNR data during the years 2000\,--\,2012. We searched for associations amongst the 134 radio events in our list with interplanetary transient phenomena detected in-situ at 1\,AU, such as shockwaves and ICMEs reported by existing catalogs. Following temporal and spatial considerations, it was possible to associate 57\% of the kmTII events to a shockwave and 46\% to an ICME. 
Of particular importance is the finding of 45 kmTII events out of of the total 134 (33\%) which had not been reported by the official Wind/WAVES catalog (based on RAD1 and RAD2 data). 

We analyzed the solar cycle variation of all kmTIIs, with associated shock and ICMEs. We find that, for the maximum of solar cycle 23, the amount of shock-associated kmTIIs is notably larger than half of the total kmTIIs, while for the other epochs this proportion is lower. KmTIIs associated with both shocks and ICMEs represent typically less than half of the total KmTIIs, except for the year 2002. 
During the years 2008\,--\,2010 no kmTIIs were found, in accordance with the yearly number of shocks which significantly decreases during those years. This is however not the case for the amount of ICMEs detected near Earth, which notably starts increasing from 2008 onwards (see Figure~\ref{fig:shocks_hist}).

ICMEs with (kmTII-ICMEs) and without (other-ICMEs) associated KmTIIs were compared in regards to their main characteristics. Regarding the mean speed of ICMEs, kmTII-ICMEs are 23\% faster on average than the other-ICMEs. This is consistent with the fact that low frequency radio bursts are associated with faster CMEs on average \citep{Patel2022}. As for the fraction of ICMEs classified as MCs, it is roughly similar for both groups of ICMEs. However, the proportion of ICMEs with BDEs is noticeably different, with 74\% of the KmTII-ICMEs group having BDEs vs. 48\% for the other-ICMEs group. Note that the presence of BDEs suggests that electrons are mirroring within closed magnetic structures with both ends anchored at the Sun's surface or disconnected plasmoids \citep{gosling90, carcaboso}.

The analysis of DsT values for both ICME groups yields that, for the kmTII-ICMEs the mean value is twice as large than for the other-ICMEs group (-111 vs. -53). This implies that kmTIIs-ICMEs produce intense storms on average, as opposite to the other-ICMEs, which tend to produce moderate storms. When considering events with Dst $<$\,-50, the storms produced by kmTII-ICMEs are 25\% larger than those from other-ICMEs (-133 vs. -102). Moreover, the proportion of ICMEs producing moderate to intense geomagnetic storms is twice as large for the kmTII-ICMEs group than for the other-ICMEs group (79\% vs. 38\%).

With regards to the differences between the number of kmTIIs compiled in our list and  Wind's list, it is worth noting that the latter reports some events to reach frequencies lower than 300 kHz that could not be identified in the TNR dynamic spectra during our survey. Some of them were even associated to structures detected in-situ at Earth. One of the possible reasons to this issue might have been that the missing kmTII events are hidden behind other radio structures in the DS, especially for solar maximum times. On the other hand, it is also plausible that the lower resolution of RAD1 at the bottom end of its frequency range may lead to misleading inferences. 

We report here on 45 new kmTII events not reported by the official Wind/WAVES catalog, which is built on the basis of RAD1 and RAD2 data. While about a third of the 45 new kmTIIs do have an apparent higher frequency counterpart, the remaining events do not. This finding poses new open questions on these low frequency TII radio bursts, regarding their generation mechanisms and the circumstances that favour their detection. It is plausible either that these new events are primarily composed of bursts that only started emitting below the frequency range of RAD1, or that they have traveled relatively unencumbered until meeting some density structure in interplanetary space. Another explanation to this finding is that kmTII emissions are wider-beamed than shorter-wavelength emissions, thus enabling their detection even if the main propagation direction of the shock is far off the Sun-Earth line.

Given that kmTIIs are closely related to the presence of shockwaves in the IP medium, it is expected that most of the kmTIIs identified in this work can be succesfully associated to a shockwave. However, this was not the case for 42$\%$ of the kmTIIs. This can be explained, at least partially, by the following:
\begin{itemize}
    \item The shock has indeed arrived but for some reason it was not detected and/or cataloged as such. The increase in the quantities that define the presence of a shock may have smoothed or somewhat diffused in the background solar wind and is too weak to be detected at 1\,AU. 
    \item In spite of the kmTII being detected by Wind/WAVES, the shock is not precisely  travelling in the direction of the spacecraft, i.e. the portion of the shockwave that produced the kmTII is not on the Sun-Earth line.
\end{itemize}

Inspired by the fact that for some of the kmTIIs lacking a reported shock in the inspected catalogs, it was still possible to infer the presence of a shock in the DS, we took a closer look at some events. In particular, for the ICMEs that produced kmTIIs but do not have associated shocks according to the inspected catalogs, we searched for shocks directly in the data. Those 11 events (Table \ref{table:resumen}) were analyzed on a case by case basis, considering magnetic field, flow speed, proton density and flow pressure. After this inquiry, we found evidence of a shock matching the ICME date of arrival for at least 9 of those 11 events. Only for 2 events we did not find any evidence of a shock. This suggests that a shock might have been responsible of the kmTII generation in most of the cases, but its properties have somewhat subsided. Therefore, it is plausible that smoothing/diffusion processes might be the reason that the ratio of kmTII associated shocks against kmTII emissions is not constant over the solar cycle.

The results of this investigation reinforce the importance of using data from TNR to identify kmTII events propagating in the interplanetary medium that cannot be detected by the other receivers of Wind/WAVES. In fact, from the 45 newly reported kmTII events, 23 are associated to an in-situ shockwave. Considering that these low-frequency radio emissions can be used as a predictor for shock arrival time at Earth with an average error of $\approx 4$\,h \citep{cremades15}, the amount of events for which space weather forecasts can be improved is increased.
\\


%
\begin{acks}
FM is fellow of CONICET. HC and FML are members of the Carrera del Investigador Cient\'ifico (CONICET). TNC acknowledges that this work falls within the objectives of LASSOS. The authors thank E. Aguilar-Rodríguez for his disinterested support on SolarSoft routines for DS generation. The authors acknowledge the use of data from the Wind (NASA) mission, as well as in situ data courtesy of N. Ness at Bartol Research Institute; D.J. McComas at SWRI; J.H. King and N. Papatashvilli at AdnetSystems NASA GSFC, and CDAWeb.

\end{acks}

%
\appendix\section{Complete table of identified events}\label{appendix}

\fontsize{7}{11}\selectfont

\begin{landscape}

\setlength{\tabcolsep}{3pt}

\begin{center}
\begin{longtable}{ccccccccccccccccccc}
\caption{List of the 134 identified kmTII events, their associations with shocks and ICMEs, and main characteristics. Columns (1-2): kmTII start date and time. (3-4) kmTII end date and time. (5) kmTII frequency range as measured in TNR's DS [kHz]. Please note that TNR's maximum frequency of observation is 256 kHz, so some kmTII may continue up to higher frequencies in the other receivers. (6) Year and reference number for the detection made by Wind/WAVES list. (7-8) Shock start date and time. (9) Shock speed [km s$^{-1}$]. (10) Shock catalog. (11-12) ICME start date and time. (13) MC type, according to RC list (see Section \ref{sec:assoc_inter} for details). (14) Magnetic obstacle type, according to TNC catalog (see Section \ref{sec:assoc_inter}). (15) Existence of bidirectional suprathermal electron strahls. (16) Maximum solar wind speed during the passage of the ICME [km s$^{-1}$]. (17) Mean magnetic field value during the ICME interval [nT]. (18) Minimum DsT value during the ICME passage [nT]. (19) ICME catalogs for which an association was found. All times are in UTC.}
\label{table:associations}\\
\cmidrule(lr){1-19}
\multicolumn{5}{c}{KmTII} & Wind/WAVES& \multicolumn{4}{c}{In situ shock} &\multicolumn{8}{c}{ICME} & \\\cmidrule(lr){1-5}\cmidrule(lr){7-10}\cmidrule(lr){11-19}

\multicolumn{2}{c}{Start date \& time} &\multicolumn{2}{c}{End date \& time} &Freq. range&Year/Ref&\multicolumn{2}{c}{Start date \& time}&Speed&Catalog&\multicolumn{2}{c}{Start date \& time}&MC&MO&BDE&V$_{ICME}$&B&DsT&Catalog \\\cmidrule(lr){1-2}\cmidrule(lr){3-4}\cmidrule(lr){7-8}\cmidrule(lr){11-12}

\multicolumn{1}{c}{(1)} &\multicolumn{1}{c}{(2)} &\multicolumn{1}{c}{(3)} &\multicolumn{1}{c}{(4)}  &(5) &(6)&\multicolumn{1}{c}{(7)} &\multicolumn{1}{c}{(8)}  &(9)&(10)&\multicolumn{1}{c}{(11)} &\multicolumn{1}{c}{(12)}  &(13) &(14)&(15)&(16)&(17) &(18) &(19) \\
\cmidrule(lr){1-19}

\endfirsthead

\multicolumn{18}{c}%
{{\tablename\ \thetable{} -- continued from previous page}} \\
\hline\multicolumn{5}{c}{Km Type II} & Wind/WAVES& \multicolumn{4}{c}{In situ shock} &\multicolumn{8}{c}{ICME} & \\\cmidrule(lr){1-5}\cmidrule(lr){7-10}\cmidrule(lr){11-19}

\multicolumn{2}{c}{Start date \& time} &\multicolumn{2}{c}{End date \& time} &Freq. range&Year/Ref&\multicolumn{2}{c}{Start date \& time}&Speed&Catalog&\multicolumn{2}{c}{Start date \& time}&MC&MO&BDE&V$_{ICME}$&B&DsT&Catalog \\\cmidrule(lr){1-2}\cmidrule(lr){3-4}\cmidrule(lr){7-8}\cmidrule(lr){11-12}

\multicolumn{1}{c}{(1)} &\multicolumn{1}{c}{(2)} &\multicolumn{1}{c}{(3)} &\multicolumn{1}{c}{(4)}  &(5) &(6)&\multicolumn{1}{c}{(7)} &\multicolumn{1}{c}{(8)}  &(9)&(10)&\multicolumn{1}{c}{(11)} &\multicolumn{1}{c}{(12)}  &(13) &(14)&(15)&(16)&(17) &(18) &(19)\\
\cmidrule{1-19}

\endhead
\hline

\endfoot

\endlastfoot

20000109 &1500 &20000110 &1800 &245-58 &2000/1 &20000111 &1340 &514.4 &ace & & & & & & & & & \\
20000129 &0630 &20000129 &1930 &200-40 &2000/5 &20000130 &1844 &642.9 &ace & & & & & & & & & \\
20000208 &2000 &20000210 &0400 &200-27 &2000/7 & & & & & & & & & & & & & \\
20000210 &1800 &20000211 &0000 &47-37 & &20000211 &0212 &549.9 &wind/ace &20000211 &0258 &0 & &Y &420 &7 &-25 &RC \\
20000211 &0830 &20000211 &2330 &55-25 &2000/9 &20000211 &2334 &638.3 &wind/ace &20000211 &2352 &2 &Fr &Y &540 &13 &-133 &RC/TNC \\
20000310 &0030 &20000312 &0300 &166-33 &2000/15 & & & & & & & & & & & & & \\
20000317 &0600 &20000318 &0030 &77-25 &2000/16 & & & & &20000318 &2200 &0 & &Y &380 &9 &-3 &RC \\
20000406 &0000 &20000406 &1300 &240-100 &2000/22 &20000406 &1632 &641.5 &wind &20000406 &1639 &1 & &Y &560 &6 &-288 &RC \\
20000516 &0440 &20000516 &1400 &166-50 &2000/38 & & & & &20000516 &2300 &1 & &N &550 &9 &-92 &RC \\
20000607 &0200 &20000607 &1300 &245-50 &2000/41 &20000608 &0841 &868.7 &ace &20000608 &0910 &0 &E &Y &610 &11 &-90 &RC/TNC \\
20000714 &1300 &20000715 &0600 &200-30 &2000/57 & & & & &20000715 &1437 &2 &F+ &Y &740 &20 &-301 &RC/TNC \\
20000716 &1300 &20000716 &2200 &245-100 &2000/59 &20000719 &1530 &638.2 &wind &20000719 &1527 &0 &Cx &Y &530 &8 &-93 &RC/TNC \\
20000726 &0900 &20000727 &0400 &200-66 &2000/66 &20000728 &0542 &474.7 &ace &20000728 &0634 &2 &F+ &Y &440 &9 &-71 &RC/TNC \\
20000905 &0300 &20000905 &1100 &90-50 &2000/67 &20000906 &1613 &539.9 &ace &20000906 &1702 & &Cx & &444 &8.93 &-35 &TNC \\
20000913 &0700 &20000913 &1230 &200-58 &2000/68 & & & & & & & & & & & & & \\
20000914 &0845 &20000914 &2130 &77-38 & &20000915 &0427 &374.7 &wind & & & & & & & & & \\
20000930 &1300 &20001001 &2130 &200-50 &2000/74 &20001003 &0102 & &wind &20001003 &0054 &2 &F+ &Y &400 &14 &-143 &RC/TNC \\
20001015 &1200 &20001016 &2300 &100-37 &2000/77 & & & & & & & & & & & & & \\
20001029 &0600 &20001030 &0300 &200-60 &2000/81 &20001031 &1630 &480.5 &ace & & & & & & & & & \\
20001101 &0330 &20001102 &1200 &83-58 &2000/82 &20001102 &2347 &279.0 &UNH & & & & & & & & & \\
20001110 &0015 &20001110 &1500 &240-55 &2000/85 &20001111 &0412 &964.8 &wind/ace &20001111 &0411 & &E &N &790 &7 &-52 &TNC \\
20001124 &1100 &20001124 &1500 &200-100 &2000/93 & & & & & & & & & & & & & \\
20001124 &2100 &20001126 &0800 &240-28 & &20001126 &1143 &524.4 &wind/ace &20001126 &1158 &0 &Fr &Y &560 &10 &-80 &RC/TNC \\
20010110 &0500 &20010111 &1300 &0-25 &2001/2 & & & & & & & & & & & & & \\
20010112 &0000 &20010113 &0000 &77-32 &2001/3 &20010113 &0225 &398.0 &wind/ace & & & & & & & & & \\
20010115 &1300 &20010115 &1930 &200-66 &2001/4 &20010117 &1622 &422.6 &wind & & & & & & & & & \\
20010118 &2100 &20010119 &0300 &100-62 &2001/5 & & & & & & & & & & & & & \\
20010122 &1700 &20010123 &0545 &59-35 & &20010123 &1049 &615.3 &wind/ace &20010123 &1048 &1 &Cx &Y &400 &4 &-61 &RC/TNC \\
20010130 &0000 &20010130 &1800 &77-35 & &20010131 &0835 &486.5 &wind/ace & & & & & & & & & \\
20010329 &1800 &20010329 &2230 &240-60 &2001/17 &20010330 &2151 &357.1 &ace & & & & & & & & & \\
20010331 &0200 &20010401 &1730 &200-43 & & & & & & & & & & & & & & \\
20010407 &0630 &20010407 &1715 &240-0 &2001/22 &20010407 &1756 &604.3 &wind/ace & & & & & & & & & \\
20010409 &1930 &20010410 &0040 &200-83 &2001/24 & & & & & & & & & & & & & \\
20010410 &1200 &20010411 &0600 &200-33 & &20010411 &1314 &560.5 &ace &20010411 &1343 &2 &F- &Y &640 &14 &-271 &RC/TNC \\
20010416 &0000 &20010416 &0630 &245-100 &2001/27 &20010418 &0049 &601.2 &wind &20010418 &0046 &0 & &Y &430 &8 &-114 & \\
20010418 &0900 &20010418 &1420 &200-111 &2001/28 & & & & & & & & & & & & & \\
20010424 &1700 &20010425 &0630 &66-40 &2001/29 & & & & & & & & & & & & & \\
20010427 &0300 &20010428 &0500 &200-20 &2001/30 &20010428 &0500 &930.1 &wind &20010428 &0501 &2 &Cx &N &550 &8 &-47 &RC/TNC \\
20010504 &0530 &20010504 &0730 &125-100 &2001/31 &20010506 &0906 &347.2 &wind & & & & & & & & & \\
20010611 &1100 &20010612 &0700 &200-59 &2001/34 & & & & & & & & & & & & & \\
20010615 &0630 &20010615 &1015 &220-58 &2001/35 & & & & & & & & & & & & & \\
20010618 &1800 &20010619 &1800 &245-71 &2001/39 & & & & & & & & & & & & & \\
20010720 &0140 &20010720 &0630 &166-71 &2001/40 & & & & & & & & & & & & & \\
20010810 &1400 &20010810 &1930 &125-83 &2001/41 &20010812 &1109 &419.1 &wind/ace & & & & & & & & & \\
20010816 &1500 &20010816 &2230 &240-66 &2001/42 &20010817 &1101 &519.1 &wind/ace &20010817 &1103 &0 &F- &Y &500 &11 &-105 &RC/TNC \\
20010825 &2100 &20010826 &1100 &240-100 &2001/43 & & & & & & & & & & & & & \\
20010912 &1000 &20010912 &1530 &142-77 &2001/46 &20010913 &0231 &449.3 &wind &20010913 &0231 &1 & &Y &410 &10 &-57 &RC \\
20010913 &1700 &20010914 &0330 &200-110 &2001/47 & & & & & & & & & & & & & \\
20010924 &1230 &20010925 &2000 &245-30 &2001/54 &20010925 &2017 &850.6 &wind &20010925 &2016 & &F- & & 639&14.12 &-102 &TNC \\
20010927 &1400 &20010927 &2300 &245-90 &2001/55 &20010929 &0929 &725.1 &wind &20010929 &0940 &1 &Cx &Y &560 &12 &-66 &RC/TNC \\
20011010 &1115 &20011010 &2200 &143-50 &2001/61 &20011011 &1650 &579.0 &wind/ace &20011011 &1701 &1 & &Y &560 &22 &-71 &RC \\
20011020 &0600 &20011021 &1600 &240-33 &2001/63 &20011021 &1640 &636.0 &wind/ace &20011021 &1648 &0 &Fr &Y &460 &9 &-187 &RC/TNC \\
20011024 &0300 &20011024 &2045 &62-33 & &20011025 &0859 &366.9 &wind/ace & & & & & & & & & \\
20011026 &0000 &20011027 &2330 &240-30 &2001/65 &20011028 &0313 &589.1 &wind &20011028 &0319 &0 &E &N &360 &5 &-157 &RC/TNC \\
20011028 &1700 &20011028 &2215 &166-100 &2001/66 & & & & & & & & & & & & & \\
20011104 &2100 &20011106 &0100 &245-35 & &20011106 &0125 & &UNH &20011106 &0152 &1 & &Y &600 &7 &-292 &RC \\
20011107 &1900 &20011108 &0800 &166-50 & & & & & & & & & & & & & & \\
20011117 &1600 &20011118 &0400 &200-66 &2001/72 &20011119 &1815 &628.9 &wind/ace &20011119 &1815 &1 & &Y &430 &6 &-47 &RC \\
20011123 &0530 &20011124 &0400 &245-30 &2001/75 &20011124 &0454 &658.7 &wind &20011124 &0656 &2 &Fr &Y &720 &14 &-221 &RC/TNC \\
20011226 &1100 &20011227 &0300 &240-90 &2001/78 &20011229 &0517 &527.4 &wind/ace &20011229 &0538 &1 &Fr &N &400 &16 &-58 &RC/TNC \\
20020214 &0600 &20020214 &1400 &245-200 &2002/5 &20020217 &0124 &300.9 &wind/ace & & & & & & & & & \\
20020311 &1030 &20020314 &1000 &181-62 & & & & & & & & & & & & & & \\
20020412 &0400 &20020415 &2300 &91-43 & & & & & & & & & & & & & & \\
20020416 &1500 &20020416 &2300 &142-83 & &20020417 &1101 &516.8 &wind/ace &20020417 &1107 &2 &F+ &Y &480 &14 &-127 &RC/TNC \\
20020417 &1330 &20020419 &0130 &240-50 &2002/17 &20020419 &0825 &768.4 &wind/ace &20020419 &0835 &2 &Fr &Y &500 &8 &-149 &RC/TNC \\
20020421 &0800 &20020421 &2400 &240-43 &2002/19 &20020423 &0500 &644.2 &wind/ace & & & & & & & & & \\
20020508 &0115 &20020508 &1845 &100-43 & &20020510 &1029 &431.3 &ace &20020510 &1109 & &Fr & &353 &6.84 &-14 &TNC \\
20020517 &0800 &20020518 &1000 &125-42 & &20020518 &1945 &545.0 &wind/ace &20020518 &1945 & &Fr & &455 &11.20 &-58 &TNC \\
20020522 &1600 &20020523 &1000 &200-0 &2002/25 &20020523 &1015 &770.6 &wind/ace &20020523 &1050 &2 &F- &Y &590 &11 &-109 &RC/TNC \\
20020524 &2300 &20020525 &1730 &143-45 & & & & & & & & & & & & & & \\
20020716 &0330 &20020716 &1600 &240-40 &2002/29 &20020717 &1555 &483.2 &wind &20020717 &1603 &1 &Fr &Y &460 &6 &-17 &RC/TNC \\
20020730 &0200 &20020801 &1200 &166-45 & &20020801 &0424 &446.5 &ace &20020801 &0510 &2 &Fr &N &450 &12 &-51 &RC/TNC \\
20020731 &1200 &20020801 &2300 &125-83 & &20020801 &2309 &495.5 &wind &20020801 &2309 &2 &Fr &Y &460 &10 &-102 &RC/TNC \\
20020817 &1130 &20020817 &2000 &220-50 &2002/41 &20020818 &1840 &671.5 &wind &20020818 &1846 &1 &Fr &Y &460 &8 &-106 &RC/TNC \\
20020824 &0700 &20020824 &1240 &250-71 & &20020826 &1041 &400.0 &UNH & & & & & & & & & \\
20020906 &0000 &20020907 &0800 &250-31 &2002/43 &20020907 &1622 &897.0 &wind/ace &20020907 &1636 &0 & &Y &470 &11 &-181 &RC \\
20021115 &0410 &20021115 &1830 &230-59 & &20021116 &2304 &514.3 &ace &20021116 &2305 &1 &F- &Y &380 &10 &-52 &RC/TNC \\
20021219 &1000 &20021220 &1700 &100-42 & & & & & &20021220 &1700 &0 &Fr &N &440 &11 &-75 &RC/TNC \\
20030201 &0230 &20030201 &2030 &245-30 & & & & & & & & & & & & & & \\
20030319 &2330 &20030320 &0930 &100-35 & & & & & & & & & & & & & & \\
20030528 &1000 &20030529 &0100 &245-50 &2003/14 &20030529 &1831 &906.6 &ace &20030529 &1825 &1 & &Y &600 &20 &-144 &RC \\
20031023 &0100 &20031023 &23:59 &170-71 & & & & & &20031024 &1524 &1 & &Y &560 &21 &-36 &RC \\
20031028 &0600 &20031029 &0600 &38-21 &2003/26 & & & & &20031029 &0611 &2 &Cx &Y &1300 &32 &-353 &RC/TNC \\
20031103 &0430 &20031103 &1930 &200-43 & &20031104 &0646 &759.0 &wind & & & & & & & & & \\
20031104 &2230 &20031105 &1800 &240-38 & & & & & & & & & & & & & & \\
20040725 &2330 &20040726 &2200 &180-22 &2004/15 &20040726 &2225 &1086.3 &wind/ace &20040726 &2249 &2 &Cx &Y &870 &16 &-170 &RC/TNC \\
20040729 &1930 &20040730 &1815 &250-35 &2004/18 &20040730 &2031 &573.7 &ace & & & & & & & & & \\
20040812 &0315 &20040812 &0930 &200-100 & & & & & & & & & & & & & & \\
20040913 &0000 &20040913 &1930 &140-35 &2004/21 & & & & &20040913 &2003 &1 &E &Y &550 &6 &-38 &RC/TNC \\
20041105 &0230 &20041105 &2300 &66-31 & & & & & & & & & & & & & & \\
20041107 &2200 &20041108 &1930 &240-30 &2004/30 &20041109 &1825 &812.9 &wind &20041109 &1825 &2 &F+ &SEP &640 &14 &-374 &RC/TNC \\
20041203 &0700 &20041204 &9999 &180-100 &2004/36 & & & & & & & & & & & & & \\
20050116 &0300 &20050117 &0600 &250-40 &2005/4 &20050117 &0715 &651.8 &ace & & & & & & & & & \\
20050117 &1300 &20050118 &1400 &250-43 &2005/6 & & & & &20050118 &2100 &0 &F- &Y &800 &12 &-80 &RC/TNC \\
20050212 &1400 &20050213 &0900 &180-60 & & & & & & & & & & & & & & \\
20050513 &2130 &20050515 &0210 &200-27 &2005/18 &20050515 &0210 &857.5 &wind &20050515 &0238 &2 &F+ &Y &630 &15 &-247 &RC/TNC \\
20050723 &1700 &20050724 &1400 &240-111 & & & & & & & & & & & & & & \\
20050724 &1530 &20050724 &2200 &240-140 &2005/29 & & & & & & & & & & & & & \\
20050730 &1130 &20050801 &0100 &250-0 &2005/32 &20050801 &0600 &479.6 &wind & & & & & & & & & \\
20050822 &0800 &20050824 &0530 &250-38 &2005/36 &20050824 &0535 &566.3 &wind/ace &20050824 &0613 &1 & &Y &660 &20 &-216 &RC \\
20050909 &2100 &20050910 &0900 &240-90 &2005/44 &20050911 &0057 &1146.6 &wind &20050911 &0114 &0 & &Y &900 &10 &-139 &RC \\
20050912 &2200 &20050913 &0615 &90-43 & & & & & &20050913 &0900 &0 & &Y &630 &5 &-86 &RC \\
20050913 &1200 &20050913 &1930 &166-83 &2005/48 & & & & & & & & & & & & & \\
20050914 &0430 &20050915 &0515 &200-34 & &20050915 &0836/0839 &663.9 &wind/ace &20050915 &0600 &1 & &Y &680 &7 &-80 &RC \\
20050915 &1700 &20050916 &0830 &125-25 &2005/50 & & & & & & & & & & & & & \\
20060817 &0030 &20060817 &0500 &200-140 & &20060818 &1548 &497.8 &wind & & & & & & & & & \\
20061202 &1350 &20061202 &2400 &200-77 & & & & & & & & & & & & & & \\
20061206 &2130 &20061207 &0700 &250-25 &2006/10 & & & & & & & & & & & & & \\
20061213 &0820 &20061213 &2340 &250-70 &2006/11 &20061214 &1351 &1011.9 &wind/ace &20061214 &1414 &2 &F- &Y &740 &13 &-162 &RC/TNC \\
20070125 &0013 &20070125 &2330 &250-70 &2007/1 & & & & & & & & & & & & & \\
20110415 &1930 &20110417 &2100 &200-59 & &20110418 &0546 &369.6 &wind & & & & & & & & & \\
20110530 &0400 &20110530 &1000 &200-100 &2011/14 & & & & & & & & & & & & & \\
20110616 &0130 &20110617 &0200 &66-20 & & & & & &20110617 &0241 &1 &Cx &Y &500 &9 &-5 &RC/TNC \\
20110804 &0700 &20110805 &1730 &240-26 &2011/23 &20110805 &1732 &411.7 &wind &20110805 &1751 &1 & &Y &540 &4 &-115 &RC \\
20110809 &2230 &20110810 &2300 &245-30 & & & & & & & & & & & & & & \\
20110922 &1600 &20110924 &1800 &245-30 &2011/31 &20110925 &1046 &265.6 &wind & & & & & & & & & \\
20111022 &1800 &20111023 &0600 &200-40 &2011/40 & & & & & & & & & & & & & \\
20111126 &1200 &20111128 &0500 &145-35 &2011/43 &20111128 &2100 &454.8 &wind &20111128 &2150 &2 &Cx &Y &450 &15 &-25 &RC/TNC \\
20120119 &1900 &20120121 &9999 &0 &2012/2 &20120121 &0402 &325.4 &wind &20120121 &0501 &2 &Fr &N &320 &10 &-10 &RC/TNC \\
20120121 &1700 &20120121 &2400 &55-35 & &20120122 &0533 &442.8 &wind &20120122 &0611 &2 &F- &Y &450 &9 &-71 &RC/TNC \\
20120123 &0700 &20120124 &1430 &245-20 &2012/3 &20120124 &1440 &735.8 &wind & & & & & & & & & \\
20120127 &2200 &20120128 &0800 &245-65 &2012/4 &20120130 &1543 &356.5 &wind & & & & & & & & & \\
20120224 &2330 &20120226 &1200 &245-30 &2012/6 & & & & &20120226 &2139 &2 &Cx &Y &440 &13 &-57 &RC/TNC \\
20120307 &0300 &20120309 &0800 &245-30 &2012/11 & & & & & & & & & & & & & \\
20120310 &2300 &20120311 &1100 &245-50 &2012/13 & & & & & & & & & & & & & \\
20120314 &0500 &20120314 &1200 &240-27 & &20120315 &1230 &650.0 &UNH &20120315 &1306 &1 &Fr &N &680 &9 &-88 &RC/TNC \\
20120420 &2115 &20120420 &2400 &245-58 & & & & & & & & & & & & & & \\
20120517 &0530 &20120517 &1930 &245-80 & & & & & & & & & & & & & & \\
20120708 &1900 &20120710 &2030 &245-35 & & & & & & & & & & & & & & \\
20120710 &1530 &20120712 &1600 &165-40 & & & & & & & & & & & & & & \\
20120717 &2345 &20120718 &0545 &200-100 &2012/39 &20120720 &0430 &440.0 &UNH & & & & & & & & & \\
20120901 &1500 &20120902 &2200 &100-30 & &20120903 &1121 &428.6 &wind &20120903 &1213 &0 & &- &430 & &-61 &RC \\
20120920 &0200 &20120920 &0530 &200-142 &2012/49 & & & & & & & & & & & & & \\
20120928 &1400 &20120929 &1300 &240-45 & &20120930 &1014 &338.2 &wind &20120930 &1131 &0 &Cx &- &310 &8 &-39 &RC/TNC \\
\bottomrule

\end{longtable}
\normalsize
\end{center}

\end{landscape}

\begin{authorcontribution}
All authors contributed to the study conception and design. Material preparation, data collection and analysis were performed by FM under the guidance of HC and FML. The first draft of the manuscript was written by FM and all authors commented on previous versions of the manuscript. All authors read and approved the final manuscript.
\end{authorcontribution}
\begin{fundinginformation}
This work is supported by projects PIP11220200102710CO (CONICET) and MSTCAME0008181TC (UTN).
\end{fundinginformation}
\begin{dataavailability}
The datasets generated and analyzed during the current study are available from the corresponding author on reasonable request. The KmTII list is available at this link: \url{https://sites.google.com/um.edu.ar/gehme/science/km-tii-catalogue}.
\end{dataavailability}
%
\begin{conflict}
The authors declare that they have no conflicts of interest.
\end{conflict}

%
%
\bibliographystyle{spr-mp-sola}
\bibliography{bibliography}  
%
%
%
%

\end{article} 
\end{document}